\documentstyle[12pt]{article}

\begin{document}

\baselineskip=7mm
\renewcommand{\arraystretch}{1.3}

\newcommand{\cf}{{ f}}
\newcommand{\TeV}{\,{\rm TeV}}
\newcommand{\GeV}{\,{\rm GeV}}
\newcommand{\MeV}{\,{\rm MeV}}
\newcommand{\keV}{\,{\rm keV}}
\newcommand{\eV}{\,{\rm eV}}
\newcommand{\Tr}{{\rm Tr}\!}
\newcommand{\be}{\begin{equation}}
\newcommand{\ee}{\end{equation}}
\newcommand{\bea}{\begin{eqnarray}}
\newcommand{\eea}{\end{eqnarray}}
\newcommand{\ba}{\begin{array}}
\newcommand{\ea}{\end{array}}
\newcommand{\bmat}{\left(\ba}
\newcommand{\emat}{\ea\right)}
\newcommand{\refs}[1]{(\ref{#1})}
\newcommand{\ler}{\stackrel{\scriptstyle <}{\scriptstyle\sim}}
\newcommand{\ger}{\stackrel{\scriptstyle >}{\scriptstyle\sim}}
\newcommand{\lag}{\langle}
\newcommand{\rag}{\rangle}
\newcommand{\ns}{\normalsize}

\begin{titlepage}
\title{{\Large\bf Cosmology of 
Radiatively Generated  Axion Scale}\\
                          \vspace{-3.5cm}
                          \hfill{\ns KAIST-TH 1/97\\}
                          \vspace{3.5cm} }

\author{Kiwoon Choi\\[.5cm]
  {\ns\it Department of Physics, Korea Advanced Institute of Science
           and Technology}\\
  {\ns\it Taejon 305-701,  Korea} \\[.5cm]}
\date{}
\maketitle
\begin{abstract} 
\baselineskip=7.2mm  {\ns}
We discuss some cosmological aspects of
supersymmetric axion models
in which the axion scale is radiatively generated
in terms of the weak scale and the Planck scale.
They include  thermal inflation,
axions produced by the decay of oscillating Peccei-Quinn flatons,
late time baryogenesis, and finally  the possibility 
to raise up the cosmological upper bound on the axion scale
in thermal inflation scenario.
\end{abstract}

\thispagestyle{empty}
\end{titlepage}

\section{Introduction}

The axion solution to the strong CP problem [1] 
involves  an intermediate scale for the spontaneous
breaking of $U(1)_{PQ}$
which is far above the electroweak scale
but still far below the Planck scale $M_P$.
It would be interesting   that 
this intermediate axion scale
appears  as a dynamical  consequence of the 
electoweak scale
and the Planck scale.
This indeed happens [2] in some class of 
spontaneously broken supergravity (SUGRA) models.
In this scheme, the early universe
experiences the so-called thermal inflation
and subsequently a period 
dominated by  coherently oscillating flaton fields [3] 
which break $U(1)_{PQ}$ spontaneously.
Here we wish to discuss some cosmological
implications of such  flatons [4].
They include 
axions produced by oscillating flatons,
late time baryogenesis, 
and finally the possibility 
of raising up the cosmological upper bound
on the axion scale in thermal inflation scenario.

\section{Radiatively Generated Axion Scale}

As an example, let us consider a variant of the model of Ref. [2] with 
superpotential  
\begin{equation}
 W = k{\phi_1^{n+2} \phi_2 \over M_P^n } + 
     h_N N N \phi_1  + \cdots 
\end{equation}
where 
$N$ is the right-handed neutrino superfield and the ellipsis
denotes  the part depending upon
the fields in the supersymmetric standard model (SSM) sector.
Here  $\phi_1$ and $\phi_2$ correspond
to flat directions when nonrenormalizable interactions and
supersymmetry breaking effects are ignored.
Including the  radiative effects of 
the strong Yukawa coupling
$h_NNN\phi_1$, the  soft mass-squared of 
$\phi_1$  becomes  {\it negative} at scales around $F_a\simeq \langle \phi_1
\rangle$,
and thereby driving $\phi_1$ to develop vacuum expectation value (VEV) at
an intermediate scale.
Neglecting the field $\phi_2$, the renormalization group improved
scalar potential for $\phi_1$  
is given by
\begin{equation}
  V = V_0 - m_1^2 |\phi_1|^2 + k^2 {|\phi_1|^{2n+4} \over M_P^{2n} }\,,
\end{equation}
where $m_1^2$ is positive and of order $m_{3/2}^2$,
and  $V_0$ is a constant of order $m_{3/2}^2F_a^2$ which is introduced
to make $V(\langle\phi_1\rangle)=0$. 
(Here we assume that SUSY is broken by  hidden sector dynamics
yielding the electroweak scale value of $m_{3/2}\simeq 10^2\sim
10^3$ GeV.)
Clearly the minimum of this scalar potential breaks $U(1)_{PQ}$ 
by
\begin{equation}
\langle \phi_1\rangle \simeq F_a \simeq (m_{3/2}M_P^{n})^{1/n+1},
\end{equation}
where we have ignored the coefficients of order unity.
Note that the integer $n$ which determines the size of $F_a$
is determined by the Peccei-Quinn (PQ) charge assignment of the model.

\section{Thermal Inflation}

The above radiative 
mechanism generating  the axion scale has substantial 
influence on the history of the universe [3].  At high temperature,
$\phi_1$  receives a 
thermal mass $\delta m_1^2\simeq  |h_N|^2T^2\gg m_1^2$ leading to 
$\langle \phi_1 \rangle =0 $.
This thermal  mass is  generated by  right-handed neutrinos
in the thermal bath.
During this period, $\langle \phi_2\rangle=0$ also. When the temperature
falls below $T \simeq V_0^{1/4}$, which is about 
$\sqrt{m_{3/2} F_a}$, the universe is dominated by
the  vacuum energy density $V_0$
and thus there appears a short period of thermal inflation.  
Below $T < m_1 \simeq m_{3/2}$,  the effective mass of $\phi_1$ becomes
negative and then  $\phi_1$ develops an intermediate scale
VEV  given by  Eq. (3). With  $\langle \phi_1\rangle \simeq
F_a$, the other flaton field
$\phi_2$ develops also a VEV of order $F_a$
through  the $A$-type soft SUSY
breaking term, $kA\phi_1^{n+2}\phi_2/M_P^n$, in the scalar potential.
This procedure makes the thermal inflation end and subsequently
the early  universe experiences a period 
dominated by
coherently oscillating PQ flaton field which corresponds to
a linear combination of $\phi_1$ and $\phi_2$ orthogonal to the axion
field.

After the period of coherent oscillation, the universe would be reheated by
the decay products of the oscillating PQ flaton $\varphi$.
The reheat temperature $T_{RH}$
is given by
\begin{equation}
 T_{RH} \simeq   g_{_{RH}}^{-1/4}\sqrt{M_P \Gamma_{\varphi} } \simeq  
\left(\frac{N_{\rm eff}}{10}\right)^{1/2} 
\left( 10^{12} {\rm GeV} \over F_a \right) 
  \left( M_{\varphi} \over 300 {\rm GeV} \right)^{3/2} {\rm GeV} \,,
\end{equation}
where $g_{_{RH}}\equiv g_*(T_{_{RH}})$ 
counts the effective number of relativistic
degree of freedom at $T_{RH}$, 
$M_{\varphi}$ denotes the flaton mass,
and we
parameterize  the  width of the flaton
decay  into thermalizable particles as 
$\Gamma_{\varphi}=N_{\rm eff}M_{\varphi}^3/64\pi F_a^2$
with $N_{\rm eff}$
presumed to be of order 
$10\sim 10^{2}$.

\section{Axion Energy Density at Nucleosynthesis}

A feature peculiar to the PQ flaton is that
its decay products include axions.
Requring 
that these axions do not spoil 
the big-bang nucleosynthesis (NS), we found [4]
\begin{equation}
\frac{B_a}{1-B_a} \leq 0.24 \, \left( \delta N_{\nu}\over 1.5 \right) 
             \left(g_{_{RH}}\over 43/4\right)^{1/3},
\end{equation}
where $B_a$ denotes the {\it effective} branching ratio 
measuring how large fraction of flatons are 
converted into axions during the reheating,
and $\delta N_{\nu}$ is the number of allowed extra neutrino
species,   which is presumed 
to be in the range $0.1 \sim 1.5$.
This  indicates that 
we  need to tune  the effective branching ratio
$B_a$ to be less than $0.02\sim 1/3$.

As is well known, generic axion models can be classified by two classes:
hadronic axion models and Dine-Fischler-Srednicki-Zhitnitskii (DFSZ)
axion models [1].
In hadronic axion models, all fields in the supersymmetric standard
model (SSM) sector carry 
{\it vanishing} PQ charge. As a result, 
flaton   couplings to SSM fields do vanish at tree level
but they appear to be nonzero by radiative effects.
Since flaton couplings to SSM fields are loop-suppressed,
in hadronic type models, most of the oscillating flatons
decay first into either axion pairs, or lighter flaton pairs, or 
flatino pairs, as long as the decays are
kinematically allowed. Lighter flatons would experience similar
decay modes, while flatinos decay into axion plus a lighter flatino.
Then in the first round of reheating,
most of flatons are converted into either axions
or the lightest flatinos. The lightest flatinos will eventually
decay into SSM particles. 
Then more than half of the original flatons
are expected to be converted into axions, i.e. the effective branching ratio
$B_a\geq 1/2$,  {\it unless} 
the flaton coupling to the  {\it lightest} flatino is unusually large.
This is in conflict with the NS limit (5) even for
the most conservative choice $\delta N_{\nu}=1.5$,
implying that hadronic axion models with  radiatively
generated axion scale
have a difficulty with the big-bang NS 
{\it unless} the models  are tuned to 
have an unusually large flaton coupling to the lightest flatino.

In  DFSZ type models, flatons have  tree level couplings to SSM
fields which are of order $M_{_{\rm SSM}}/F_a$
or  $M_{_{\rm SSM}}^2/F_a$
where $M_{_{\rm SSM}}$ collectively denotes 
the mass parameters in the SSM, e.g. $M_t$, $M_W$, $\mu$, $A$,
and so on. 
If $M_{\varphi}\gg M_{_{\rm SSM}}$,
the reheating procedure would be similar to 
that of hadronic axion models, which is problematic.
However if $M_{\varphi}$ is comparable to $M_{_{\rm SSM}}$,
the NS limit (5) does  not provide any meaningful
restriction on DFSZ type models.

\section{Late Time Baryogenesis}

Thermal inflation driven by PQ flatons is expected to dilute away any
pre-existing baryon asymmetry.  However, PQ flatons themselves  can 
produce baryon asymmetry after thermal inflation.
A careful examination of the flaton couplings in DFSZ
type models
suggests  that,  among the decays into SSM particles,
the decay channels to the top ($t$) and/or stop ($\tilde{t}$)
pairs are most important.
Stops produced by the oscillating flatons
would be in out-of equilibrium and   subsequently 
experience a $B$ and $CP$ violating decay 
to generate a baryon asymmetry 
provided that the $B$-violating operator, e.g., $\lambda''_{332}
U^c_3 D^c_3 D^c_2$ and the corresponding complex trilinear soft-term
are present [5].  Note that the PQ symmetry 
can be arranged so that
dangerous lepton-number violating operators $LQD^c, LLE^c$ are 
forbidden for the proton stability.

In order for the baryon asymmetry (generated as above)
not to be erased,
the reheat temperature (4) has  to be less than few GeVs
[5]. This means that the above mechanism for baryogenesis
can work only for $n=2$ or 3 [see Eqs.~(3) and (4)].
The produced baryon asymmetry is [4] 
\begin{equation}
 {\eta \over 3\times10^{-10}} \simeq 
|\lambda''_{332}|^2
\left({\rm arg}(Am_{1/2}^*)\over 10^{-2}\right)
\left(10^{14} {\rm GeV} \over F_a \right)
 \left( M_\varphi \over 300 {\rm GeV} \right)^{1/2},
\end{equation}
where  ${\rm arg}(A m_{1/2}^*)$ denotes the CP violating
relative phase which is constrained to be less than
about $10^{-2}$ for superparticle masses of order 100
GeV.

\section{Raising up the Upper Bound on $F_a$}

For a successful nucleosynthesis,
we need $T_{RH} > 6$ MeV 
[6],
implying  that only $n=1$, 2, and 3 are allowed.
As is well known, 
the axion scale is constrained by  the 
consideration of the coherent axion energy density produced by an initial
misalignment [1]. 
If there is no entropy production after the axion start
to oscillate at around
$T\simeq 1$ GeV, this lead to the usual bound:
$F_a\leq 10^{12}$ GeV.
When $n=2$ or 3, 
the corresponding axion
scale $F_a\simeq (M_{\varphi}M_P^n)^{1/n+1}$ would exceed this bound.
However in this case,
the reheat temperature (4) goes  
below 1 GeV.
Then the coherent axions may be significantly diluted by 
the entropy dumped from flaton decays, thereby 
allowing  $F_a$ much bigger than $10^{12}$ GeV [7].
Taking into account of this dilution,
we find [4]
\begin{equation}
\Omega_a h_{50}^2\simeq  
\left(N_{\rm eff}\over 10\right)
\left( 10^{12} {\rm GeV} \over F_a \right)^{0.44}
\left(\frac{M_{\varphi}}{300 \, {\rm GeV}}\right)^{2.9}
\left( \Lambda_{QCD} \over 200 \, {\rm MeV} \right)^{-1.9},
\end{equation}
where we have used 
$\Gamma_{\varphi}\simeq N_{\rm eff}M_{\varphi}^3/64\pi F_a^2$.
The above result  is valid only for $n\geq 2$.
As we have anticipated, it shows that the case of $n=2$ or 3
with $F_a\simeq (M_{\varphi}M_P^n)^{1/n+1}$
yields  a coherent axion
energy density not exceeding the critical density although
the corresponding $F_a$ exceeds $10^{12}$ GeV.
Furthermore, in this case of $n=2$ or 3,
{\it axions can be a good dark matter
candidate} for an appropriate value of $M_{\varphi}$, which
was  not possible for $n=1$.

\end{document}